\documentclass[a4paper,11pt]{article}
\pdfoutput=1
\usepackage{multirow}

\usepackage{jcappub} 

\usepackage{appendix}
\usepackage{multirow}

\title{\boldmath{IDECAMB: an implementation of interacting dark energy cosmology in CAMB}}

\author[a]{Yun-He Li}
\author[a,b,c,d,1]{and Xin Zhang\note{Corresponding author.}}
\affiliation[a]{Department of Physics, College of Sciences, Northeastern University, Shenyang 110819, China}
  \affiliation[b]{Key Laboratory of Cosmology and Astrophysics (Liaoning), Northeastern University, Shenyang 110819, China}
\affiliation[c]{Key Laboratory of Data Analytics and Optimization for Smart Industry (Ministry of Education), Northeastern University, Shenyang 110819, China}
\affiliation[d]{National Frontiers Science Center for Industrial Intelligence and Systems Optimization, Northeastern University, Shenyang 110819, China}
\emailAdd{liyunhe@mail.neu.edu.cn}
\emailAdd{zhangxin@mail.neu.edu.cn}

\abstract{Interacting dark energy (IDE) scenario is a natural and important extension to the standard $\Lambda$CDM cosmology. We develop a full numerical routine, called {\tt IDECAMB}, as a patch to the public Einstein-Boltzmann solver {\tt CAMB}, to solve the background and perturbation equations of the IDE models. The {\tt IDECAMB} solver provides a unified interface for the widely studied IDE models by employing a parametrization model with five free functions. By configuring these five functions, one can easily map the coupled quintessence (CQ) and coupled fluid (CF) models into the parametrization. We handle the perturbation evolutions of the CF models with the parametrized post-Friedmann (PPF) approach to avoid the possible large-scale instability. Compared with the previous established PPF approach whose form depends on a specific IDE model, the PPF approach in this work are model-independent, making it easy to use. We constrain a specific CQ model with the {\tt IDECAMB} package. The fitting results are consistent with those obtained by Planck Collaboration, which confirms the validity of the package.}

\begin{document}
\maketitle
\flushbottom

\section{Introduction}\label{sec:intro}

The late-time Universe is dominated by dark sectors, namely, dark matter and dark energy, which is supported by recent observations \cite{Planck:2018vyg,BOSS:2016wmc,Pan-STARRS1:2017jku}. Dark matter provides the force for the structure formation, while dark energy accounts for the cosmic acceleration \cite{SupernovaSearchTeam:1998fmf,SupernovaCosmologyProject:1998vns}. The primary candidate of dark energy is the cosmological constant $\Lambda$ with the equation of state (EoS) equal to $-1$. The corresponding $\Lambda$ cold dark matter ($\Lambda$CDM) model can fit the cosmological data with the least parameters, and is commonly viewed as the standard cosmological model. However, the so-called fine-tuning and coincidence problems \cite{Weinberg:1988cp,Sahni:1999gb,Peebles:2002gy,Bean:2005ru,Copeland:2006wr,Sahni:2006pa,Kamionkowski:2007wv,Frieman:2008sn,Li:2011sd} always suffer this model on the theoretical level. Besides, it is also reported recently that some discrepancies exist in the model between some measurements \cite{DiValentino:2021izs,Verde:2019ivm,DiValentino:2017gzb,DiValentino:2020zio,Freedman:2017yms,Riess:2019qba,Heymans:2020gsg}. This situation prompts cosmologists to consider some complex models, such as the dynamical dark energy with an EoS not exactly equal to $-1$, or the modified gravity (MG) theory trying to explain the cosmic acceleration outside the framework of General Relativity (see ref. \cite{Weinberg:2013agg} for a recent review). 

In standard cosmology, dark energy and dark matter are generally assumed to be non-interacting. This means that they do not exert any influence on each other, except through gravitational interaction. However, from the view of particle physics, interactions are ubiquitous in the world. Thus, unless forbidden by a fundamental symmetry, a direct interaction between the dark sectors is a natural way to extend the $\Lambda$CDM cosmology. In such an interacting dark energy (IDE) scenario, the conservation laws for the energy-momentum tensor ($T_{\mu\nu}$) of dark energy ($de$) and cold dark matter ($c$) are modified, 
\begin{equation}
  \label{eq:energyexchange} \nabla_\nu T^\nu_{\hphantom{j}\mu,de} = -\nabla_\nu T^\nu_{\hphantom{j}\mu,c}  = Q_\mu,
\end{equation} 
where $Q_\mu$ denotes the energy-momentum transfer vector. A specific $Q_{\mu}$ carries the information of three important interaction quantities: the energy transfer rate $Q$, the energy transfer perturbation $\delta Q$ and the momentum transfer rate $f$. Due to them, the background and perturbation evolutions of the dark sectors are modified, which not only provides a possible mechanism to alleviate the coincidence problem \cite{Amendola:1999er,Comelli:2003cv,Zhang:2005rg,Cai:2004dk}, but also introduces new features to the structure formation \cite{Amendola:2001rc,Bertolami:2007zm,Koyama:2009gd}.

Plenty of IDE models has been proposed \cite{Amendola:1999er,He:2008tn,He:2009mz,He:2009pd,He:2010im,Boehmer:2008av,Guo:2007zk,Xia:2009zzb,Wei:2010cs,Li:2011ga,Zhang:2004gc,Li:2013bya,Wang:2013qy,Pourtsidou:2013nha,Yang:2017ccc,Yang:2018euj,Kase:2020hst,Linton:2021cgd,Faraoni:2014vra,Zhang:2012uu,Zhang:2013lea,Valiviita:2008iv}, among which two types of the IDE models are widely studied. One is the so-called coupled quintessence (CQ) model, in which dark energy is a quintessence field $\phi$. The coupling between the quintessence field and the dark matter field is defined by a Lagrangian, giving $Q_\mu \propto\partial_\mu\phi$. The corresponding $Q$, $\delta Q$ and $f$ can be obtained with a few assumptions in the CQ model. On the other hand, if dark energy is described by some fluid model, there is no a fundamental theory pointing out how $Q_{\mu}$ can be constructed. In practice, one often assumes it to be proportional to the four velocity ($u_{\mu}$) of the dark fluid, namely, $Q_{\mu}\propto u_{\mu,I}$ for $I=c$ or $de$. The energy transfer rate $Q$ in such model is also constructed, phenomenologically, such as letting $Q\propto \rho_{c}$, $Q\propto \rho_{de}$ or proportional to some linear or nonlinear combinations of the energy densities ($\rho$) of the dark sectors. For convenience, we call them the coupled fluid (CF) models.

In this study, our primary focus is on the interacting dark energy (IDE) scenario. Our main objective is to develop a convenient and efficient method for testing IDE models with observations.
It is well-known that testing cosmological models with observations is an important mission in cosmology.
To do this, we need to solve the full dynamic equations of the models and then confront their predictions with the observational data. The public {\tt CAMB} and {\tt CosmoMC} packages \cite{Lewis:1999bs,Lewis:2002ah} provide us the powerful tools to do this, but only natively support a few non-interacting dark energy models. To adapt for an IDE model, one needs to carefully modify the background and the perturbation systems of the {\tt CAMB} solver. For the various IDE models if we implement this one by one, it is however cumbersome. Perhaps, a more efficient way is trying to establish a unified solver for the IDE theory by using a parametrization technology. Then one can test the IDE theory in a model-independent way, or constrain a specific model by simply mapping it into the parametrization. The widely used {\tt MGCAMB} \cite{Zhao:2008bn,Hojjati:2011ix} and {\tt EFTCAMB} \cite{Hu:2013twa,Hu:2014sea} packages are the successful examples of this strategy implemented for the MG theory.

However, before doing this for IDE models, one should particularly note the large-scale instability problem \cite{Valiviita:2008iv} widespread in the CF models ($Q_{\mu}\propto u_{\mu,I}$ for $I=c$ or $de$). The cosmological perturbations will blow up on super-horizon scales if the EoS of dark energy and the coupling constant take some specific values (for more details, see refs. \cite{Valiviita:2008iv,He:2008si,Clemson:2011an}). To avoid this instability, one has to carefully set the priors for the parameters in the observational constraints, which is not only inconvenient but also unable to reflect the actual preference of the data sets \cite{Li:2014cee}. It can be concluded that the power of a unified solver for IDE models will be greatly reduced, if this problem is not well handled. Fortunately, we have a solution to this issue. In ref.~\cite{Li:2014eha}, we introduce the parametrized post-Friedmann (PPF) approach \cite{Hu:2008zd,Fang:2008sn} into the calculations of the dark energy perturbation in the IDE scenario, for the first times. We find that the large-scale instability can be successfully avoided, and the full parameter space of the CF models can be explored with the extended PPF approach.

So now, we have the opportunity to develop a fully functional solver for IDE models, which is exactly what we aim to accomplish in this work. We will employ a parametrization model to establish a unified interface for the two types of the IDE models. Due to the fact that the CF model cannot be defined by a fundamental theory, as mentioned above, it is also hard to construct a parametrization form of $Q_{\mu}$ from a unified Lagrangian. As an alternative we follow ref.~\cite{Skordis:2015yra} and directly parametrize the two perturbation quantities $\delta Q$ and $f$, which can be written as the linear combinations of the perturbations of the dark sectors. We will rewrite the form of the PPF approach for the parametrized IDE model. The PPF approach established in previous works~\cite{Li:2014eha,Li:2014cee} depends on a specific IDE model, making it hard to use for other researchers. In this paper, with the parametrized IDE model, we can establish a model-independent form of the PPF approach, so one can use it without concern for the details of the PPF approach. Finally, we develop a full numerical routine to solve the background and perturbation equations of the IDE models. We shall call the code {\tt IDECAMB}\footnote{Downloaded from \url{https://github.com/liaocrane/IDECAMB/}.}, which can be used as a patch to the {\tt CAMB} and {\tt CosmoMC} packages to test the IDE theory.

Our paper is organized as follows. In section \ref{sec:GeneralEqs}, we give the general equations in the IDE theory. In section \ref{sec:Model}, we briefly review the CQ and the CF models, and then construct a parametrized IDE model for them. The standard linear perturbation theory as well as the PPF approach for the perturbation evolutions of this parametrized IDE model are given in section \ref{sec:evolution}. The background and the perturbation equations used in the numerical calculations are shown in section \ref{sec:numericalEqs}. In section \ref{sec:fitting} we show an example of the observational constraint on a specific IDE model. The conclusions of this paper are given in section \ref{sec:conclusions}. Some necessary calculations are shown in appendixes \ref{app:Coupledquintessence} and \ref{app:PPF_S}.

\section{General equations}\label{sec:GeneralEqs}
We start from the standard linear perturbation theory. Using the eigenfunctions of the Laplace operator, $Y$, and its covariant derivatives, $Y_i=-k^{-1}\nabla_i Y$, and $Y_{ij}=(k^{-2}\nabla_i\nabla_j+\gamma_{ij}/3)Y$ with $-k^2$ the eigenvalue of the Laplace operator and $\gamma_{ij}$ the spatial metric, the Friedmann-Robertson-Walker (FRW) metric with scalar perturbations can be expressed in general as~\cite{Kodama:1984ziu,Bardeen:1980kt}
\begin{align}
   &  {g_{00}} = -a^{2} (1+2 AY),\qquad {g_{0i}} = -a^{2} B Y_i,  \nonumber \\
   & \qquad {g_{ij}} = a^{2} (\gamma_{ij}+2 H_L Y \gamma_{ij} + 2 {H_T Y_{ij}}),
  \label{eqn:metric}
\end{align}
where $a$ is the scale factor of the Universe, and the four functions, $A$, $B$, $H_L$, and $H_T$, denote the amplitudes of four types of metric perturbations.
Similarly, the energy-momentum tensor can also be expressed as
\begin{align}
   & {T^0_{\hphantom{0}0}} =  - (\rho+ \delta\rho Y),\qquad{T^i_{\hphantom{0}0}} = -(\rho + p){v}Y^i, \nonumber \\
   & \qquad  {T^i_{\hphantom{i}j}} = (p+{\delta p}Y)  \delta^i_{\hphantom{j}j}
  + p{\Pi Y^i_{\hphantom{j}j}},
  \label{eqn:dstressenergy}
\end{align}
where $p$ denotes the pressure, and the four functions, $\delta\rho$, $v$, $\delta p$, and $\Pi$ represent the energy density perturbation, velocity, isotropic pressure perturbation, and anisotropic stress, respectively.

The energy-momentum transfer vector $Q_{\mu}$ in eq.~(\ref{eq:energyexchange}) can be split in general as \cite{Kodama:1984ziu},
\begin{equation}
  Q_{\mu}  = a\Big( -Q(1+AY) - \delta QY,\,[ f+ Q (v-B)]Y_i\Big).\label{eq:Qenergy}
\end{equation}
Then, eq.~(\ref{eq:energyexchange}) in the background level reduces to
\begin{gather}
    \rho'_{de} +3\mathcal{H}(1+w)\rho_{de}= aQ,\label{eq:rhodedot}\\
    \qquad\rho'_{c} +3\mathcal{H}\rho_{c}=- aQ,\label{eq:rhocdot}
\end{gather}
where $w={p_{de}/\rho_{de}}$ is the EoS of dark energy, $\mathcal{H}=a'/a$ is the conformal Hubble parameter, and a prime denotes the derivative with respect to the conformal time. In the linear perturbation level, eq.~(\ref{eq:energyexchange}) gives
\begin{gather}
   {\delta\rho_I'}+  3\mathcal{H}({\delta \rho_I}+ {\delta p_I})+(\rho_I+p_I)(k{v}_I + 3 H_L')=a(\delta Q_I+AQ_I),\label{eqn:conservation1}                                                                 \\
   [(\rho_I + p_I)\theta_I]'+4\mathcal{H}(\rho_I + p_I)\theta_I -{ \delta p_I } +{2 \over 3}c_K p_I {\Pi_I} - (\rho_I+ p_I) {A}=a(Q_I\theta+f_{k,I}),\label{eqn:conservation2}
\end{gather}
where $c_K = 1-3K/k^2$ with $K$ the spatial curvature, and for convenience we define 
\begin{equation}
  f_{k,I}={f_I\over k},\quad\theta_I={v_I-B\over k}. \label{eq:fk_theta}
\end{equation}
Note that in our notation, $Q_{de}=-Q_{c}=Q$, $\delta Q_{de}=-\delta Q_{c}=\delta Q$, and $f_{de}=-f_{c}=f$, indicating that the directions of the energy and momentum transfer are from cold dark matter to dark energy for a positive $Q$, $\delta Q$, and $f$.

\section{Models and parametrization}\label{sec:Model}
The coupling of the dark sectors is completely described by the three quantities: $Q$, $\delta Q$ and $f_k$. In this section, we first show the specific forms of these three quantities for some typical CQ and CF models. 
Next, we delve into the construction of parameterization forms for $\delta Q$ and $f_k$, aiming to capture the essential characteristics of the two types of models.

\subsection{Coupled quintessence models}
The CQ models are defined by the following Lagrangian,
\begin{equation}
  {\cal L} =
  -\frac{1}{2}\partial^\mu \phi \partial_\mu \phi - U(\phi) -
  m(\phi)\bar{\psi}\psi + {\cal L}_{{\rm kin},\psi},\label{eq:L_cq}
\end{equation}
where dark energy is a quintessence field $\phi$ with a potential $U(\phi)$, and the mass $m$ of the matter field $\psi$ depends on the value of $\phi$ due to their interaction. Considering a non-universal coupling, namely, the case of $\phi$ only interacting with cold dark matter field, then we can obtain the energy-momentum transfer vector as 
\begin{equation}\label{eq:covQ_CQ}
  Q_{ \mu} = \frac{\partial
    \ln{m(\phi)}}{\partial \phi} \rho_c \, \partial_\mu \phi .
\end{equation}
Obviously, once a specific $m(\phi)$ is given, the interaction forms in the background and perturbation levels can be derived. In practice, an exponential coupling is widely studied in the literature \cite{Pettorino:2008ez,Pettorino:2013oxa,Amendola:2002bs,Pettorino:2012ts,Amendola:2011ie,Planck:2015bue}, namely
\begin{equation}
  m(\phi)\propto e^{-\beta\sqrt{\kappa}\phi},
\end{equation} 
with $\kappa=8\pi G$ and $\beta$ the coupling constant. Substituting this coupling into eq. (\ref{eq:covQ_CQ}), we get the energy-momentum transfer vector, 
\begin{equation}\label{eq:covQ_CQ_model}
  Q_\mu=-\beta\rho_c\sqrt{\kappa}\partial_\mu\phi.
\end{equation} 

Comparing eq. (\ref{eq:covQ_CQ_model}) with eq. (\ref{eq:Qenergy}), one can get the background energy transfer,
\begin{equation}
  Q=\beta\rho_c{\sqrt{\kappa}\phi'\over a}, \label{eq:Q_CQ}
\end{equation}
the energy transfer perturbation, 
\begin{equation}\label{eq:covQ_CQ_dQ}
  \delta Q=Q\Big({\delta \phi '\over \phi'}+{\delta \rho_c \over \rho_c}-A\Big),
\end{equation}
and the momentum transfer potential, 
\begin{equation}\label{eq:covQ_CQ_f}
  f_k=-Q\theta+Q{\delta \phi \over \phi'}.
\end{equation}

\subsection{Coupled fluid models} 
For the CF models, there is no fundamental theory to point out how $Q_{\mu}$ can be constructed. In practice, one often assumes it to be proportional to the four velocity of the dark fluid, namely  
\begin{equation}\label{eq:covQ_CF}
  Q_{\mu}=Q u_{\mu,I},
\end{equation}
where
\begin{equation}
  u_{\mu,I} = a\Big(-1-AY,\,(v_I-B)Y_i \Big),
\end{equation}
with $I=c$ or $de$. In such a construction, the energy momentum transfer is vanished in the rest frame of $I$ fluid. The energy transfer rate $Q$ in the CF model is also constructed, phenomenologically, such as letting $Q=\beta H\rho_{de}$ or $Q=\beta H\rho_{c}$, where the Hubble parameter $H=\mathcal{H}/a$ is introduced to equalize the dimensions. One can also let $Q=\beta H_0\rho_{de}$ or $Q=\beta H_0\rho_{c}$ to obtain an $H$-independent model \cite{Valiviita:2008iv}, where $H_0$ is the Hubble constant.

Comparing with eq. (\ref{eq:Qenergy}), one can find that the momentum transfer potential in the CF model is 
\begin{equation}\label{eq:covQ_CF_f}
  f=Q(v_I-v).
\end{equation} 
The energy transfer perturbation $\delta Q$ can be obtained by directly perturbing $Q$ in the linear order. For example, we have
\begin{equation} 
\delta Q=Q\Big({\delta \rho_{de}\over \rho_{de}}+{\delta \rho_c \over \rho_c}\Big),
\end{equation}
for the non-linear model $Q=\beta H \rho_{de}\rho_c/(\rho_{de}+\rho_c)$ (originated from the generalized Chaplygin gas model \cite{Zhang:2004gc,Li:2013bya,Wang:2013qy}). Here we do not consider the perturbation of the Hubble parameter, but note that $\delta H$ is indispensable if one tries to get the gauge invariant equations in an $H$-dependent model \cite{Gavela:2010tm}.

\subsection{Parametrization model}
As mentioned above, we cannot construct a parametrization form of $Q_{\mu}$ from a unified Lagrangian. As an alternative, we directly parametrize $\delta Q$ and $f_k$. It is not hard to find that $\delta Q$ and $f_k$ in the CF models are generally linear combinations of $\delta_{de}$, $\delta_c$, $\theta_{de}$ and $\theta_c$, where $\delta_{I}=\delta\rho_I/\rho_I$. In fact, in the CQ models, $\delta Q$ and $f_k$ can also be written as the functions of these four quantities. Treating quintessence as a dark energy fluid, we can get the relations between the field perturbations $\delta\phi$ and $\delta\phi'$, and the fluid perturbations $\theta_{de}$ and $\delta\rho_{de}$. See this process in appendix \ref{app:Coupledquintessence}. Then, using eqs.~(\ref{eq:drhode_quint}) and (\ref{eq:vpide_quint}), we can rewrite eqs.~(\ref{eq:covQ_CQ_dQ}) and (\ref{eq:covQ_CQ_f}) as, 
\begin{gather}
  \delta Q={Q\over 1+w}\delta_{de}+Q\delta_c-a\beta\rho_c\sqrt{\kappa} U_\phi\theta_{de},\\
  f_k=-Q\theta+Q\theta_{de},
\end{gather}
where $U_{\phi }$ denotes the derivative of $U$ with respect to $\phi$.

Thus, we can use five functions $C_1$, $C_2$, $C_3$, $D_1$, and $D_2$ to parametrize the energy and momentum transfer perturbations as
\begin{align}
  \delta Q & =C_1\delta_{de}+C_2\delta_{c}+C_3\theta_{de}, \label{eq:parametrized_dQ}\\
  f_k      & =-Q\theta+D_1\theta_{de}+D_2\theta_c.\label{eq:parametrized_f}
\end{align}
{For a CF model, $C_3$ is generally zero and the remaining four functions can parametrize the common CF models. For example, considering a general CF model with $Q_\mu=\beta^{ij} H\rho_{i} u_{\mu,j}$, one can get $Q=\sum_{j=1}^{2}\beta^{ij} H\rho_{i}$, $C_1=\sum_{i=1}^{2}\beta^{1i} H\rho_{1}$, $C_2=\sum_{i=1}^{2}\beta^{2i} H\rho_{2}$, $D_1=\beta^{i1}H \rho_{i}$ and $D_2=\beta^{i2}H \rho_{i}$, where $\rho_{i}=(\rho_{de},\,\rho_{c})$, $u_{\mu,i}=(u_{\mu,de},\,u_{\mu,c})$ and each element of $\beta^{ij}$ can be a free parameter or a free function of $a$. Note that the repeated Latin letters here represent summation as usual. This model reduces to a simple linear CF model if $\beta^{ij}$ has only one non-zero element. Moreover, if choosing $\beta^{11}=-\beta^{12}$ and $\beta^{21}=-\beta^{22}$, we have $Q,\,C_1,\,C_2=0$ and hence $\delta Q=0$, corresponding to a pure momentum transfer model introduced in ref. \cite{Simpson:2010vh}. Besides the CF model and the CQ model, our parametrization can also describe the so-called Type 1 model proposed in ref.~\cite{Pourtsidou:2013nha} in which dark energy is described by a k-essence field \cite{Armendariz-Picon:2000nqq,Armendariz-Picon:2000ulo}. In table \ref{tab:modelfuncs}, we present the examples of some typical IDE models mapping into the parametrization model. Replacing $H$ with $H_0$ in the table, one can get the corresponding $H$-independent models. Our parametrization provides a unified interface for the widely studied IDE models with the least free functions. One can also find a more general parametrization in ref.~\cite{Skordis:2015yra}.}

\begin{table}[tbp]
  \centering
  \newcommand{\tabincell}[2]{\begin{tabular}{@{}#1@{}}#2\end{tabular}}
  \begin{tabular}{llccccc}
    \hline
    Model                                             & $Q$                                                          & $C_1$          & $C_2$    & $C_3$                      & $D_1$    & $D_2$    \\
    \hline
    \multirow{3}{*}{$Q_\mu=Q u_{\mu,c}$}              & $\beta H\rho_{de}$                                   & $Q$            & $\cdots$ & $\cdots$                   & $\cdots$ & $Q$      \\
                                                      & $\beta H\rho_{c}$                                    & $\cdots$       & $Q$      & $\cdots$                   & $\cdots$ & $Q$      \\
                                                      & $\beta H{\rho_{de}\rho_{c}\over \rho_{de}+\rho_{c}}$ & $Q$            & $Q$      & $\cdots$                   & $\cdots$ & $Q$      \\
    \hline
    \multirow{3}{*}{$Q_\mu=Q u_{\mu,de}$}             & $\beta H\rho_{de}$                                   & $Q$            & $\cdots$ & $\cdots$                   & $Q$      & $\cdots$ \\
                                                      & $\beta H \rho_{c}$                                    & $\cdots$       & $Q$      & $\cdots$                   & $Q$      & $\cdots$ \\
                                                      & $\beta H{\rho_{de}\rho_{c}\over \rho_{de}+\rho_{c}}$ & $Q$            & $Q$      & $\cdots$                   & $Q$      & $\cdots$ \\

    \hline
    $Q_\mu=\beta^{ij} H\rho_{i} u_{\mu,j}$      & $\sum\limits_{j=1}^{2}\beta^{ij} H\rho_{i}$  & $\sum\limits_{i=1}^{2}\beta^{1i} H\rho_{1}$  & $\sum\limits_{i=1}^{2}\beta^{2i} H\rho_{2}$ & $\cdots$ & $\beta^{i1}H \rho_{i}$      & $\beta^{i2}H \rho_{i}$ \\                
    \hline
    $Q_\mu=-\beta\rho_c\sqrt{\kappa}\partial_\mu\phi$ & $\beta\rho_c{\sqrt{\kappa}\phi'\over a}$ & ${Q\over 1+w}$ & $Q$      & ${a^2U_\phi\over -\phi'}Q$ & $Q$      & $\cdots$ \\
    \hline
  \end{tabular}
  \caption{Some typical IDE models mapping into the parametrization model. Replacing $H$ with $H_0$, one can get the corresponding $H$-independent models. The $Q_\mu=\beta^{ij} H\rho_{i} u_{\mu,j}$ model represents a general linear CF model, where $\rho_{i}=(\rho_{de},\,\rho_{c})$, $u_{\mu,i}=(u_{\mu,de},\,u_{\mu,c})$ and each element of $\beta^{ij}$ can be a free parameter or function of $a$.}\label{tab:modelfuncs}
\end{table}

\section{Perturbation evolutions of the dark sectors}\label{sec:evolution}
In this section, we handle the perturbation evolutions of the dark sectors for the parametrized IDE model, as constructed above. For cold dark matter, the dynamic systems described by eqs.~(\ref{eqn:conservation1}) and (\ref{eqn:conservation2}) can be completed by the conditions, $\delta p_c=0$ and $\Pi_c=0$. Substituting eqs.~(\ref{eq:parametrized_dQ}) and (\ref{eq:parametrized_f}) into eqs.~(\ref{eqn:conservation1}) and (\ref{eqn:conservation2}), we have 
\begin{gather}
    \delta_c'+k^2\theta_c+kB +3H_L' ={a\over\rho_c}(Q\delta_c-QA-C_1\delta_{de}-C_2\delta_c-C_3\theta_{de}),       \\
    \theta_c'+{\cal H}\theta_c-A= {a\over\rho_c}(Q\theta_c-D_1\theta_{de}-D_2\theta_c).
\end{gather}
For dark energy, besides $\Pi_{de}=0$, we still need another condition to complete the dynamic systems. In the following, we provide two methods to calculate the perturbations of dark energy, namely the standard linear perturbation theory as well as the PPF approach.

\subsection{The standard linear perturbation theory}
In the standard linear perturbation theory, the dynamic systems of dark energy are completed by adding the information of $\delta p_{de}$, whose value can be calculated in terms of a rest-frame sound speed $c_{s}$ with $c_{s}^2={\delta p_{de}\over\delta\rho_{de}}|_{\mathrm{rf}}$. Here the subscript ``rf'' denotes the dark energy rest frame ($v_{de}=0$ and $B=0$). Making a gauge transformation from the dark energy rest frame gauge to a general gauge, one can get
\begin{equation}
  \delta p_{de} = c_{s}^2\delta\rho_{de} -\rho_{de}'(c_{s}^2-c_{a}^2)\theta_{de},\label{eq:cs}
\end{equation}
where $c_{a}^2={p_{de}'\over\rho_{de}'}$ and $c_{a}$ is the adiabatic sound speed of dark energy. If dark energy is an adiabatic fluid, $c_{s}^2=c_{a}^2$, the pressure perturbation of dark energy only has the adiabatic mode, namely $\delta p_{de} = c_{a}^2\delta\rho_{de}$. However, one can find that $c_{s}^2=w<0$ (for the constant $w$ case), which will lead to a nonphysical result for dark energy collapsing~\cite{Gordon:2004ez}. Thus, in practice, dark energy is generally taken as a non-adiabatic fluid with a positive $c_{s}^2$.

Substituting eq.~(\ref{eq:cs}) into eqs.~(\ref{eqn:conservation1}) and (\ref{eqn:conservation2}), we have
\begin{align}
    \delta_{de}'&+3{\cal H}(c_{s}^2-w)\delta_{de}+(1+w)(k^2\theta_{de}+kB+3H_L')+3{\cal H}(c_{s}^2-c_{a}^2)\Big[3{\cal H}(1+w)-{aQ\over\rho_{de}}\Big]\theta_{de} \nonumber \\
                &={a\over\rho_{de}}(QA-Q\delta_{de}+C_1\delta_{de}+C_2\delta_{c}+C_3\theta_{de}), \label{eq:SLT_drhode}   \\
    \theta_{de}'&+{\cal H}(1-3c_{s}^2)\theta_{de}-{c_{s}^2\over 1+w}\delta_{de}-A = {a\over\rho_{de}(1+w)}\big[D_1\theta_{de}+D_2\theta_{c}- Q(1+c_{s}^2)\theta_{de}\big].\label{eq:SLT_vde}
\end{align}
These two equations can be used for the perturbation calculations of the CQ models, if the values of $w$, $c_{s}^2$ and $c_{a}^2$ are obtained for quintessence dark energy. In fact, treating quintessence field as a fluid and using eqs.~(\ref{eq:drhode_quint})--(\ref{eq:vpide_quint}), we can directly obtain  
\begin{align}
  \delta p_{de}=\delta\rho_{de}-2\phi'  U_{\phi }\theta_{de}.\label{eq:cs_quint}
\end{align}
Comparing eq.~(\ref{eq:cs_quint}) with eq.~(\ref{eq:cs}), one can immediately find that $c_{s}^2=1$ for quintessence field. Using eqs.~(\ref{eq:rhode_quint}) and (\ref{eq:pde_quint}) in combination with eq.~(\ref{eq:rhodedot}), we can also get
\begin{align}
  w=-1+{\phi '^2\over a^2\rho_{de}},
\end{align}
and
\begin{align}
  c_a^2=1-{2a^2\phi'U_\phi\over a^3Q-3\mathcal{H}\phi'^2}.
\end{align}
Thus, once the background evolutions for a CQ model are obtained, one can get the dark energy perturbations by directly solving eqs.~(\ref{eq:SLT_drhode}) and (\ref{eq:SLT_vde}) instead of evolving the field equations for $\delta\phi$ and $\delta\phi'$.   

On the other hand, if we directly use eqs.~(\ref{eq:SLT_drhode}) and (\ref{eq:SLT_vde}) to calculate the perturbations of dark energy in a CF model, some problems will occur. From eqs.~(\ref{eq:rhodedot}) and (\ref{eq:cs}), one can find that the energy transfer rate $Q$ will enter the non-adiabatic part of $\delta p_{de}$. For some values of $w$ and $\beta$, the non-adiabatic mode will grow fast on the large scales, leading to rapid growth of the curvature perturbation at the early times~\cite{Valiviita:2008iv}. This is the well-known large scale instability in the CF models. In fact, even for the non-interacting dark energy the calculation of $\delta p_{de}$ in eq.~(\ref{eq:cs}) can also bring instability when $w$ crosses the phantom divide $w=-1$ \cite{Vikman:2004dc,Hu:2004kh,Caldwell:2005ai,Zhao:2005vj}. Thus, we do not use eqs.~(\ref{eq:SLT_drhode}) and (\ref{eq:SLT_vde}) to handle the dark energy perturbations in the CF models. As an alternative, we calculate them by the following PPF approach.

\subsection{The PPF approach}

The PPF approach is originally designed for testing the MG theory \cite{Hu:2007pj}, while it is found to be effective to eliminate the instability when $w$ crosses the phantom divide $w=-1$ in a non-interacting dark energy model \cite{Hu:2008zd,Fang:2008sn}. Inspired by this, we introduce the PPF approach into the calculations of the perturbations in the IDE scenario in ref.~\cite{Li:2014eha}. We find that the large-scale instability problem can be successfully resolved with the extended PPF approach. In the following, we first briefly review the construction of the PPF approach, and then show how it is applied to the parametrized IDE model.

The PPF approach is established in the comoving gauge, defined by $B=v_T$ and $H_T=0$, where $v_T$ denotes the velocity of total matters except dark energy. For convenience, we use the new symbols, $\zeta\equiv H_L$, $\xi\equiv A$, $\rho\Delta\equiv\delta\rho$, $\Delta p\equiv\delta p$, $V\equiv v$, $\Theta\equiv\theta$, and $\Delta Q\equiv\delta Q$, to denote the corresponding quantities of the comoving gauge except for the two gauge independent quantities $\Pi$ and $f_k$. The PPF approach gives an approximate value of $V_{de}$ on the large scales, and compromise the perturbation evolutions on the large scales and small scales using an empirical formula. In this process, the dynamic systems of dark energy perturbations can be completed without the information of $\delta p_{de}$. 

To be specific, on the large scales ($k_H=k/{\cal H}\ll 1$), we can establish a relationship between $V_{de}-V_T$ and $V_T$. Since ($V_{de}-V_T)/V_T={\cal O}(k_H^2)$ at $k_H\ll 1$, this relationship can be exactly parametrized by a function $f_\zeta$ (see refs.~\cite{Fang:2008sn,Li:2014eha}). However, in practice, it suffices for most purposes to simply let $f_\zeta=0$ \cite{Fang:2008sn,Li:2014cee}. Thus we directly let $V_{de}=V_T$ in this paper. With this condition, the Einstein equation for $\zeta$ reduces to 
\begin{align}
  \lim_{k_H \ll 1} \zeta'  = \mathcal{H}\xi - {K \over k} V_T.
  \label{eqn:zetaprimesh}
\end{align}
On the small scales, the Poisson equation gives $\zeta+V_T/k_H=\kappa a^2\Delta_T \rho_T/( 2k^2c_K)$. The PPF approach introduces a dynamical quantity $\Gamma$ to compromise the perturbations on the large and small scales, and thus on all scales we have
\begin{equation}
  \zeta+{V_T\over k_H}+\Gamma = {\kappa a^2
  \over  2k^2c_K} \Delta_T \rho_T.
  \label{eqn:modpoiss}
\end{equation}
If the equation of motion for $\Gamma$ is obtained, the dynamic system is completed. Obviously, on the small scales, $\Gamma\rightarrow0$ at $k_H\gg1$. On the other hand, the derivative of eq.~(\ref{eqn:modpoiss}) in combination with eq.~(\ref{eqn:zetaprimesh}) and the Einstein equations gives the equation of motion for $\Gamma$ on the large scales,
\begin{equation}\label{eq:gammadot}
  \lim_{k_H \ll 1} \Gamma'  = S -\mathcal{H}\Gamma,
\end{equation}
where
\begin{equation}
  S=  {\kappa a^2
      \over 2k^2 } \Big[\rho_{de}( 1+ w)kV_T -{3\mathcal{H}a\over c_K}(Q\Theta+f_k)-\frac{a}{c_K}(\Delta Q+\xi Q)\Big],\label{eq:PPF_S}
\end{equation}
and $\xi$ can be obtained from eq.~(\ref{eqn:conservation2}) using $\Theta_T=0$,
\begin{equation}
  \xi =  -{\Delta p_T - {2\over 3}c_K p_T \Pi_T-a(Q\Theta+f_k) \over \rho_T + p_T}.
  \label{eq:PPF_xi}
\end{equation}
Then, with a transition scale parameter $c_\Gamma$ ($c_\Gamma=0.4c_s$ in practice), we can take the equation of motion for $\Gamma$ on all scales to be \cite{Hu:2008zd,Fang:2008sn}
\begin{equation}
  (1 + c_\Gamma^2 k_H^2) [\Gamma' +\mathcal{H} \Gamma + c_\Gamma^2 k_H^2 \mathcal{H}\Gamma] = S.
  \label{eq:PPF_gammaeom}
\end{equation}
Once the equation of motion for $\Gamma$ is solved, the dark energy perturbations can be obtained by
\begin{gather}
  \kappa a^{2}\rho_{de}\Delta_{de} =-2 k^{2}c_K  \Gamma- 3\kappa a^{2}\rho_{de}(1+w)\mathcal{H} \Theta_{de},\label{eq:PPF_drho_de} \\
  \kappa a^2 \rho_{de}( 1+ w)\Theta_{de}F =-2(S - \Gamma' - \mathcal{H}\Gamma) ,\label{eq:PPF_theta_de}
\end{gather}
with $F = 1 +  3 \kappa a^2 (\rho_T + p_T)/( 2k^2 c_K)$.

Now we can apply the PPF approach to the parametrized IDE model. However, if we directly substitute eqs. (\ref{eq:parametrized_dQ}) and (\ref{eq:parametrized_f}) into eqs. (\ref{eq:PPF_S}) and (\ref{eq:PPF_xi}), the equation of motion for $\Gamma$ cannot be solved, since the values of $\Delta_{de}$ and $\Theta_{de}$ are still unknown at this moment. So before solving eq.~(\ref{eq:PPF_gammaeom}), we should first strip $\Delta_{de}$ and $\Theta_{de}$ from eqs. (\ref{eq:PPF_S}) and (\ref{eq:PPF_xi}). The detailed process of this part is shown in appendix \ref{app:PPF_S}. From eqs.~(\ref{eq:PPF_S_node}) and (\ref{eq:gammadot}), we have
\begin{equation}
  S=S_0+{ aC_1\over \rho_{de}}\Gamma,
\end{equation}
where 
\begin{equation}
  S_0= {\kappa a^2
      \over 2k^2 } \Big[\rho_{de}(1+w)kV_T-\frac{a}{c_K}(3\mathcal{H}D_2\Theta_c+C_2\Delta_{c}+Q\xi_0)\Big],\label{eq:PPF_S0}
\end{equation}
with
\begin{equation}
  \xi_0 =  -{\Delta p_T - {2\over 3}c_K p_T \Pi_T-aD_2\Theta_c \over \rho_T + p_T}.
  \label{eq:PPF_xi0}
\end{equation}
Now the equation of motion for $\Gamma$ is independent of $\Delta_{de}$ and $\Theta_{de}$, and can be solved.

\section{Equations for numerical calculations }\label{sec:numericalEqs}
\subsection{The background evolutions}
We solve the background evolutions of dark sectors in the CF models with the following two differential equations,
\begin{gather}
  {d(\rho_{de}a^4)\over da}= (1-3w)\rho_{de}a^3+ {a^4Q\over\mathcal{H}},\\
  {d(\rho_{c}a^4)\over da}= \rho_{c}a^3- {a^4Q\over\mathcal{H}}.
\end{gather}
Here we evolve $\rho_{de}a^4$ and $\rho_{c}a^4$ instead of $\rho_{de}$ and $\rho_{c}$ for numerical stability. The conformal Hubble expansion rate is given by the Friedmann equation,
\begin{equation}
  \mathcal{H}=\Big[{\kappa a^2\over 3}(\rho_{de}+\rho_{c}+ ...)\Big]^{1\over2},
\end{equation}
where ``...'' represents the rest of the energy densities of the Universe. For a specific CF model with $w=w(a;\mathcal{H},\rho_{de},\rho_{c})$ and $Q=Q(a;\mathcal{H},\rho_{de},\rho_{c})$, above equations can be exactly solved using initial conditions at $a=1$. Obviously, this solver is not only suitable for a model with simple form of $w$ or $Q$, but also for some complex nonlinear models, such as the holographic dark energy (HDE) model \cite{Li:2004rb} with   
\begin{equation}
  w_{hde}=-{1\over3}-{2\over{3c}}\sqrt{\kappa a^2\rho_{de}\over3\mathcal{H}^2},
\end{equation}
where $c$ is a free parameter in the HDE model.

For the CQ models, we directly evolve the field equations (\ref{eq:fieldeq_bk}),
\begin{gather}
  {d(a^2\phi')\over da}=-{a^3\over \mathcal{H}}U_\phi+{a^4 Q\over \mathcal{H}\phi'},\\
  {d\phi\over da}={1\over \mathcal{H}a}\phi'.
\end{gather}
Given a specific CQ model with $U=U(\phi)$ and $Q=Q(a;\phi,\phi',\rho_c)$, above equations can be solved by choosing proper initial conditions at early radiation-dominated epoch. Note that $\rho_c$ is given by
\begin{equation}
  \rho_{c}=\rho_{c0}a^{-3} \exp\Big({-\int^{\phi}_{\phi_0}\beta\sqrt{\kappa} d\phi}\Big),\label{eq:rhoc_CQ}
\end{equation}
and if $\beta$ is a constant, $\rho_{c}=\rho_{c0}a^{-3} \exp[-\beta\sqrt{\kappa}(\phi-\phi_0)]$, where $\phi_0$ is the present-day value of $\phi$.

\subsection{Perturbations in the synchronous gauge}
The numerical codes for the perturbation equations are written in the synchronous gauge which is defined by $A=B=0$, $\eta=-H_T/3-H_L$, and $h=6H_L$. For cold dark matter, we have
\begin{gather}
    \delta_c'+kv_c +{h'\over2} ={a\over\rho_c}\Big(Q\delta_c-C_1\delta_{de}-C_2\delta_{c}-C_3{v_{de} \over k}\Big), \\
   v_c'+{\cal H}v_c= {a\over\rho_c}(Qv_c-D_1v_{de}-D_2v_{c}).
\end{gather}

For the perturbations of dark energy in the synchronous gauge, we also give the standard linear theory and the PPF approach. Substituting $A=B=0$ and $h=6H_L$ into eqs.~(\ref{eq:SLT_drhode}) and (\ref{eq:SLT_vde}), the standard linear theory gives
\begin{align}
    \delta_{de}'&+3{\cal H}(c_{s}^2-w)\delta_{de}+(1+w)\Big(kv_{de}+{h'\over2}\Big)+3{\cal H}(c_{s}^2-c_{a}^2)\Big[3{\cal H}(1+w)-{aQ\over\rho_{de}}\Big]{v_{de}\over k} \nonumber \\
                &={a\over\rho_{de}}\Big(-Q\delta_{de}+C_1\delta_{de}+C_2\delta_{c}+C_3{v_{de}\over k}\Big), \label{eq:SLT_drhode_sync}   \\
    v_{de}'&+{\cal H}(1-3c_{s}^2)v_{de}-{c_{s}^2\over 1+w}k\delta_{de} = {a\over\rho_{de}(1+w)}\big[D_1v_{de}+D_2v_{c}- Q(1+c_{s}^2)v_{de}\big].\label{eq:SLT_vde_sync}
\end{align}
These equations can be utilized in the perturbation calculations of the CQ models. For the CF models, to avoid the possible large-scale instability, we use the following PPF equations in the synchronous gauge. 

The connections between the comoving gauge and the synchronous gauge are \cite{Hu:2008zd}
\begin{gather}
  \rho_I\Delta_I  = \delta_I \rho_I - \rho_I' v_T/k, \label{eq:transdelta}\\
  \Delta p_I  = \delta p_I  - p_I' v_T/k, \label{eq:transdp}\\
  V_I-V_T=v_I-v_T, \label{eq:transv}\\
  \zeta = -\eta - {v_{T}/k_{H}}. \label{eq:transzeta}
  \end{gather}
Another useful transformation relation is
  \begin{equation}
  V_{T} =v_{T}+\sigma,\label{eq:transvt}
  \end{equation}
where
\begin{equation}
  {\sigma\over k_H}={\kappa a^2\over 2k^2}\Big\{{\delta\rho_T\over c_K} +{v_T\over c_Kk_H}\Big[3(\rho_T+p_T)+{aQ\over\mathcal{H}}\Big]\Big\} +\eta-\Gamma.
\end{equation}
Using above relations, we can obtain
\begin{equation}
    S_0={\kappa a^2\over 2k^2} \Big\{k\rho_{de}(1+w)(v_T+\sigma)-{3aD_2\over k_Hc_K}(v_{c}-v_T)-\frac{aQ}{c_K}\xi_0-\frac{aC_2}{c_K}\Big[\delta_c+\Big(3\mathcal{H}+{aQ\over\rho_c}\Big) {v_T\over k}\Big]\Big\},
\end{equation}
and
\begin{equation}
  \xi_0 =  {k\delta p_T-p_T'v_T - {2\over 3}kc_K p_T \Pi_T-aD_2(v_c-v_T) \over -k(\rho_T + p_T)}.
\end{equation}
Then the $\Gamma$ equation of eq.~(\ref{eq:PPF_gammaeom}) can be solved in the synchronous gauge. The corresponding dark energy perturbations are given by
\begin{gather}
  \delta_{de} =- 3\mathcal{H}(1+w){v_{de}\over k }+{aQ\over\rho_{de}}{v_T\over k }-{2k^{2}c_K \over \kappa a^{2}\rho_{de}} \Gamma,\\
  v_{de} =v_{T}-{2k( S - \Gamma' - \mathcal{H}\Gamma) \over \kappa a^2 \rho_{de}(1+w)F }.
\end{gather}
\section{Constraints on a specific model}\label{sec:fitting}

The background and perturbation equations shown in section~\ref{sec:numericalEqs} have been embedded in the public {\tt CAMB} code \cite{Lewis:1999bs}. We dub the modified code {\tt IDECAMB}. In this section, to show the reliability of the {\tt IDECAMB}, we use it in combination with the {\tt CosmoMC} package \cite{Lewis:2002ah} to constrain a specific IDE model. Actually, the {\tt IDECAMB} code for the CF models has been widely tested in our recent works \cite{Guo:2017hea,Guo:2017deu,Li:2017usw,Feng:2017usu,Guo:2018gyo,Feng:2018yew,Li:2018ydj,Feng:2019mym,Feng:2019jqa,Li:2020gtk,Zhang:2021yof,Wang:2021kxc,Zhao:2022bpd}. Thus, here we show an example of constraint on a CQ model with the energy transfer $Q$ given by eq.~(\ref{eq:Q_CQ}). For the potential of the quintessence, we choose an inverse power-law form \cite{Pettorino:2008ez}, 
\begin{equation}
  U=U_0(\sqrt{\kappa}\phi)^{-\alpha},
\end{equation}
with $U_0$ and $\alpha$ being two constants. This model has been widely constrained in the literatures \cite{Pettorino:2008ez,Pettorino:2013oxa,Amendola:2002bs,Pettorino:2012ts,Amendola:2011ie,Planck:2015bue}. In particular, Planck Collaboration obtains rich results on the free parameters $\alpha$ and $\beta$ by using their 2015 release of the cosmic microwave background (CMB) in combination with other data sets \cite{Planck:2015bue}. Note that $U_0$ and $\phi_0$ (in eq.~(\ref{eq:rhoc_CQ})) are not the free parameters, and their values must satisfy the requirement of getting correct present-day energy densities of dark energy and cold dark matter. We can use an iteration to get the true values of them in the numerical codes.

To make a direct comparison with the Planck 2015 results~\cite{Planck:2015bue}, we follow their data usage but only choose one typical data combination: TT+lowP+BSH (2015). Here TT+lowP represents the CMB high-$\ell$ TT spectrum in combination with the low-$\ell$ temperature-polarization data from Planck 2015~\cite{Planck:2015fie}, and BSH is the combination of the baryon acoustic oscillations (BAO) measurements from SDSS \cite{Ross:2014qpa}, BOSS \cite{BOSS:2013rlg}, and 6dFGS \cite{Beutler:2011hx}, the type-Ia supernovae (SNIa) data from the Joint Light-curve Analysis sample \cite{SDSS:2014iwm}, and the Hubble constant ($H_0$) measurement $H_0=(70.6\pm3.3)\,{\rm km\,s^{-1}\,Mpc^{-1}}$ \cite{Efstathiou:2013via}. The free parameters are the same as those in ref.~\cite{Planck:2015bue}, except that we take $H_0$ as a free parameter instead of the commonly used $\theta_{\rm{MC}}$, because $\theta_{\rm{MC}}$ is dependent on a standard non-interacting background evolution. We set priors [0, 1.4] for $\alpha$ and [0, 0.15] for $\beta$, and keep the priors of other parameters the same as those used by Planck Collaboration \cite{Planck:2015fie}. Since there is no large-scale instability in the CQ models, we directly use eqs.~(\ref{eq:SLT_drhode_sync}) and (\ref{eq:SLT_vde_sync}) to calculate the dark energy perturbations. The corresponding forms of the five functions $C_1$, $C_2$, $C_3$, $D_1$, and $D_2$ for the CQ model are give in table \ref{tab:modelfuncs}.

\begin{table}[tbp]
\centering
  \begin{tabular} { l c c c}
    \hline
              &  TT+lowP& TT,TE,EE+lowP & TT,TE,EE+lowE \\
    Parameter &    +BSH (2015)          & +BSH (2015)                  & +BAO+SNIa (2018)\\
   \hline
   $\alpha $     & $< 0.62 \,(95\%)$   & $< 0.57\,(95\%)       $  & $< 0.49   \,(95\%)      $\\
  
  $\beta  $     & $0.036^{+0.020}_{-0.016}   $     & $0.037^{+0.018}_{-0.014}   $& $0.036^{+0.018}_{-0.013} $\\
  
  $H_0$ (km/s/Mpc)  & $67.78^{+0.87}_{-0.77}       $ & $67.82\pm 0.79             $& $68.03\pm 0.72    $\\
  
  $\Omega_m $    & $0.3045\pm 0.0090          $    & $0.3048\pm 0.0089          $&$0.3038\pm 0.0084     $\\
  
  $\sigma_8 $     & $0.836\pm 0.020            $   & $0.840\pm 0.017            $&$0.821^{+0.012}_{-0.015} $\\
   \hline
   \end{tabular}
\caption{Marginalized mean values and 68\% C.L. intervals or the 95\% C.L. upper limits for the parameters of the CQ model.}\label{tab:params}
\end{table}

\begin{figure}[tbp]
  \centering
  \includegraphics[width=14cm]{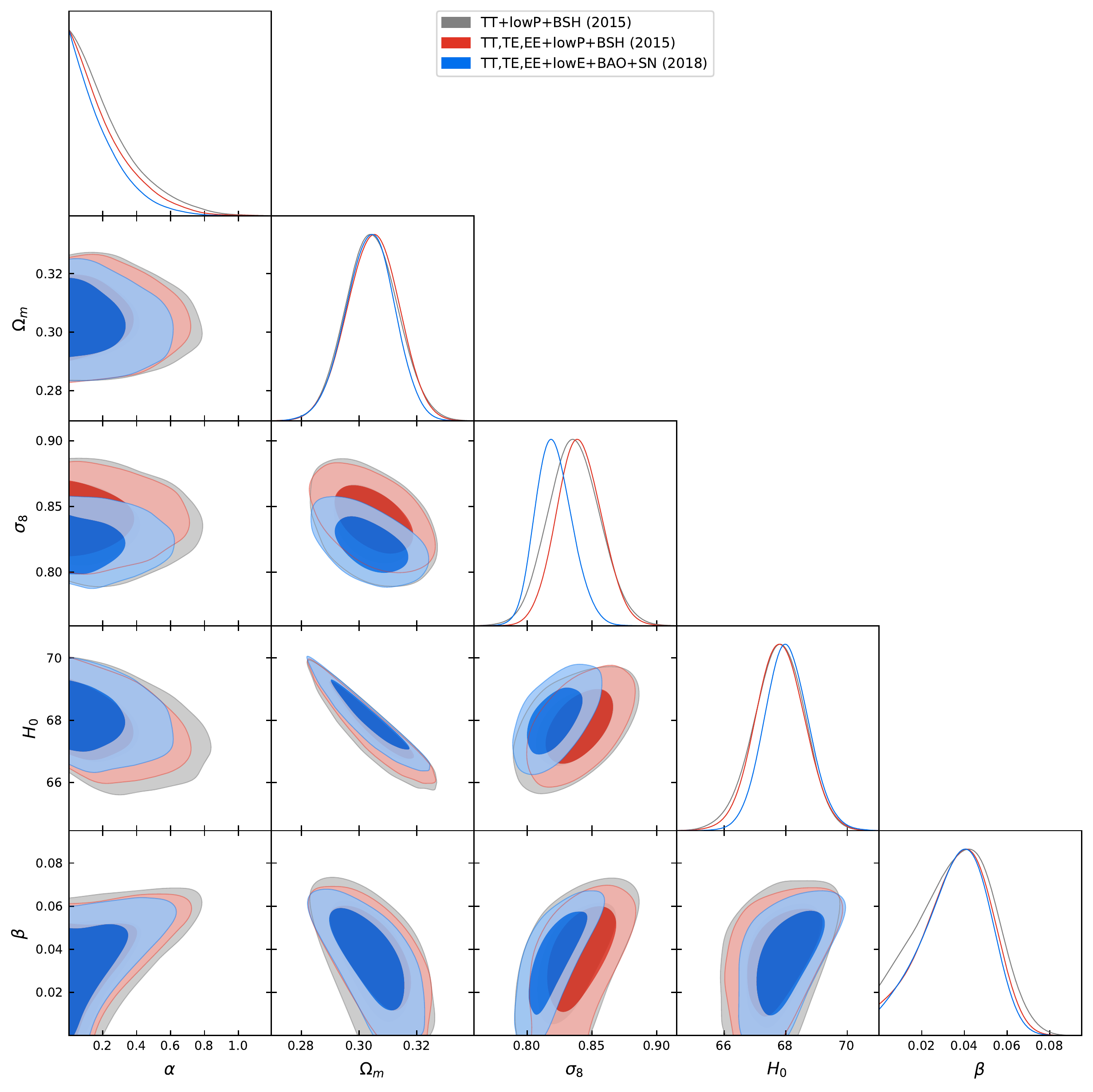}
  \caption{The one-dimensional and two-dimensional posterior distributions for the parameters of the CQ model with the TT+lowP+BSH (2015), the TT,TE,EE+lowP+BSH (2015) and the TT,TE,EE+lowE+BAO+SNIa (2018) data combinations.}\label{fig:fitresult_tri}
\end{figure}

The fit results are shown in table \ref{tab:params} and figure~\ref{fig:fitresult_tri}. Using TT+lowP+BSH (2015) data combination, we get $\beta=0.036^{+0.020}_{-0.016}$ at $1\sigma$ level, similar with the result $\beta=0.037^{+0.018}_{-0.015}$ in ref.~\cite{Planck:2015bue}. For the result of $\alpha$, Planck Collaboration obtains a non-zero value with $\alpha=0.29^{+0.077}_{-0.260}$ at $1\sigma$ level, while we only get an upper limit with $\alpha< 0.62$ at $2\sigma$ level. This difference deserves our further tests. By carefully checking our theory and codes, as well as the public information about the Planck's settings, we conclude that it mainly results from the different priors set for $\alpha$. As shown in ref. \cite{Planck:2015bue}, they have to let $\alpha \geq 0.03$ for the numerical stability in the iteration for $U_0$ and $\phi_0$, while in our iterative routine, there is no such scruple. We have tested that a similar result can be obtained in our program if setting $\alpha \geq 0.03$ by hand. 

To further show the agreement with the Planck's results, we also do a test by adding the Planck 2015 high-$\ell$ polarization (TE,EE) information into analysis. The $1\sigma$ errors or the $2\sigma$ upper limits of the parameters are slightly shrunk, as shown in table \ref{tab:params} and figure~\ref{fig:fitresult_tri}. The TT,TE,EE+lowP+BSH (2015) data sets give $\alpha< 0.57$ at $2\sigma$ level, and $\beta=0.037^{+0.018}_{-0.014}$ at $1\sigma$ level, which are in concordance with $\alpha< 0.58$ at $2\sigma$ level, and $\beta=0.036^{+0.016}_{-0.013}$ at $1\sigma$ level in ref. \cite{Planck:2015bue}. Given the fact, we believe that our program can repeat the previous results and the validity of the {\tt IDECAMB} package is confirmed.

{Having verified the reliability of the program, now we shall improve the constraints by using the latest observational data. The CMB data are updated to the Planck 2018 data release including the TT, TE, EE spectra at $\ell\geq 30$, the low-$\ell$ temperature {\tt Commander} likelihood, and the low-$\ell$ {\tt SimAll} EE likelihood~\cite{Planck:2018vyg}. For the BAO measurements, we keep the SDSS and the 6dFGS data but replace above BOSS data with the latest data release 12~\cite{BOSS:2016wmc}. For the SNIa data, we use the Pantheon sample, comprised of 1048 data points~\cite{Pan-STARRS1:2017jku}. The recent local $H_0$ measurement, such as the result of $H_0=(74.03{\pm1.42})\,{\rm km\,s^{-1}\,Mpc^{-1}}$~\cite{Riess:2019cxk}, is not included in the analysis, since it is in tension with the Planck data to a certain extent. The fit results of using this TT,TE,EE+lowE+BAO+SNIa (2018) data combination are also shown in table \ref{tab:params} and figure~\ref{fig:fitresult_tri}. 
With the new data, the 95\% C.L. upper limit of $\alpha$ is further reduced to 0.49, while there is no significant change for the value of $\beta$ compared with the result by the TT,TE,EE+lowP+BSH (2015) data combination. The marginalized posterior distribution $\beta=0.036^{+0.018}_{-0.013}$ (68\% C.L.) indicates that a non-zero coupling is preferred by the data at about 2$\sigma$. However, the goodness of fit does not point towards a deviation from the $\Lambda$CDM model at 2$\sigma$. In fact, due to the degeneracy between $\alpha$ and $\beta$, the contours in figure~\ref{fig:fitresult_tri} are almost compatible with $\Lambda$CDM ($\alpha=0$ and $\beta=0$) at 1$\sigma$. These results, from the aspect of interacting dark energy, enrich the discussions of the Planck 2018 paper \cite{Planck:2018vyg} about the extensions to the base-$\Lambda$CDM model.  }

\section{Conclusion}\label{sec:conclusions}
The IDE scenario represents a natural and significant extension to the standard $\Lambda$CDM cosmology. With numerous IDE models being proposed, there is a pressing need for efficient and rapid methods to test these models using observational data.
In this paper, we establish a unified solver for the IDE theory by using a parametrization technology. Due to the fact that the IDE models cannot be defined by a unified Lagrangian, we directly parametrize the energy transfer perturbation $\delta Q$ and the momentum transfer rate $f_k$. With five free functions $C_1$, $C_2$, $C_3$, $D_1$ and $D_2$, the parametrization forms of $\delta Q$ and $f_k$ are written as the linear combinations of the perturbations of the dark sectors. The widely studied CQ ($Q_\mu \propto\partial_\mu\phi$) and CF ($Q_{\mu}\propto u_{\mu,I}$ for $I=c$ or $de$) models can be easily mapped into this parametrization. Based on this parametrization, we develop a full numerical routine, called {\tt IDECAMB}, to solve the background and perturbation equations of the IDE models. Using the {\tt IDECAMB} solver as a patch to the public {\tt CAMB} and {\tt CosmoMC} packages, one can conveniently test the IDE models with observations. As an example, we constrain a specific CQ model with $Q_\mu=-\beta\rho_c\sqrt{\kappa}\partial_\mu\phi$ and $U=U_0(\sqrt{\kappa}\phi)^{-\alpha}$. {The fit results are consistent with those obtained by Planck Collaboration with the TT,TE,EE+lowP+BSH (2015) data sets. Using the latest TT,TE,EE+lowE+BAO+SNIa (2018) data combination, we get $\alpha< 0.49$ at $2\sigma$ level, and $\beta=0.037^{+0.018}_{-0.014}$ at $1\sigma$ level. These results confirm the validity of the {\tt IDECAMB} package.}  

A unique virtue of our {\tt IDECAMB} solver is that the well-known large-scale instability problem existing in the CF models is well handled, which benefits from the idea of calculating the perturbations of dark energy using the PPF approach instead of the standard linear perturbation theory. Although this idea was first proposed in our previous work, the form of the PPF approach established in that work depends on a specific IDE model, making it hard to modify and use for other researchers. In this paper, with the help of the parametrized IDE model, we establish a model-independent form of PPF approach, and hence people can use the {\tt IDECAMB} solver to solve their own models without concern for the details of the PPF approach. The shortage of the {\tt IDECAMB} solver is that it currently only supports the CQ model with an exponential coupling and an inverse power-law potential of the quintessence. The CQ models with other types of the couplings or potentials are not tested in this work. Moreover, in addition to the widely studied CQ and CF models, there are other types of the IDE models in the literature (see e.g. ref.~\cite{Pourtsidou:2013nha}). These models are also deserved to be included in the {\tt IDECAMB} system. We leave these in future works.

\appendix
\section{Coupled quintessence}\label{app:Coupledquintessence}
The energy-momentum tensor of the quintessence is defined by
\begin{equation}
  T_{\mu\nu}=\partial_\mu\phi\partial_\nu\phi-{1\over2}\partial_\lambda\phi\partial^\lambda\phi g_{\mu\nu}-U(\phi)g_{\mu\nu}.\label{eq:CQTensor}
\end{equation}
Using eqs.~(\ref{eqn:metric}) and $\phi=\bar{\phi}+\delta\phi Y$ (we will omit the bar in the following discussion), above equation can be expressed as
\begin{align}
  & {T^0_{\hphantom{0}0}} =-{\phi '^2\over 2a^2}-U-\left( {\phi '\over a^2} \left(\delta \phi '-A \phi '\right)+\delta \phi  U_{\phi }\right)Y,                  \\
  & {T^i_{\hphantom{0}0}} = -\frac{\phi ' }{a^2}(k\delta \phi+B\phi ') Y^i,                                                                \\
  & {T^i_{\hphantom{i}j}} =  {\phi '^2\over 2a^2}-U+\left( {\phi '\over a^2} \left(\delta \phi '-A \phi '\right)-\delta \phi  U_{\phi }\right)Y\delta^i_{\hphantom{i}j}.
 \label{eq:dstressenergy_CQ}
\end{align}
If we treat the quintessence as a fluid with the energy-momentum tensor given by eqs.~(\ref{eqn:dstressenergy}), then above equations give
\begin{gather}
  \rho_{de}={\phi '^2\over 2a^2}+U, \label{eq:rhode_quint}\\
  p_{de}={\phi '^2\over 2a^2}-U,\label{eq:pde_quint}
\end{gather}
in the background level, and 
\begin{gather}
  \delta\rho_{de}={\phi '\over a^2} \left(\delta \phi '-A \phi '\right)+\delta \phi  U_{\phi },\label{eq:drhode_quint}\\
  \delta p_{de}={\phi '\over a^2} \left(\delta \phi '-A \phi '\right)-\delta \phi  U_{\phi },\label{eq:dp_quint}\\
  \theta_{de}={v_{de}-B\over k}={\delta \phi \over\phi '},\ \Pi_{de}=0, \label{eq:vpide_quint}
\end{gather}
in the perturbation level.

The conservation law, $\nabla_\nu T^\nu_{\hphantom{j}\mu} = Q_\mu$, gives the background field equation
\begin{equation}
  \phi''+2\mathcal{H}\phi'+a^2U_\phi={a^3Q \over \phi'}.\label{eq:fieldeq_bk}
\end{equation}

\section{Equation for $\Gamma$ without $\Delta_{de}$ and $\Theta_{de}$}\label{app:PPF_S}
Substituting eqs. (\ref{eq:parametrized_dQ}) and (\ref{eq:parametrized_f}) into eqs. (\ref{eq:PPF_S}) and (\ref{eq:PPF_xi}), we have
\begin{equation}
  S=  {\kappa a^2
      \over 2k^2 } \Big[\rho_{de}( 1+ w)kV_T-{3\mathcal{H}a\over c_K}(D_1\Theta_{de}+D_2\Theta_c) - \frac{a}{c_K}(C_1\Delta_{de}+C_2\Delta_{c}+C_3\Theta_{de}+\xi Q)\Big],\label{eq:PPF_S_rw}
\end{equation}
and,
\begin{equation}
  \xi =  -{\Delta p_T - {2\over 3}c_K p_T \Pi_T-a(D_1\Theta_{de}+D_2\Theta_c) \over \rho_T + p_T}.
\end{equation}
Above equation can be rewritten as, 
\begin{equation}
  \xi =  \xi_0+{aD_1\over\rho_T + p_T}\Theta_{de},\label{eq:PPF_xi_rw}
\end{equation}
with $\xi_0$ given by eq.~(\ref{eq:PPF_xi0}).
Substituting eq.~(\ref{eq:PPF_xi_rw}) into eq.~(\ref{eq:PPF_S_rw}), we have
\begin{equation}
  S  =S_0-{\kappa a^3\over 2k^2c_K }C_1\Delta_{de}-{\kappa a^3\over 2k^2c_K }\Big(C_3+3\mathcal{H}D_1+{aQD_1\over\rho_T + p_T}\Big)\Theta_{de},
\end{equation}
where $S_0$ is given by eq.~(\ref{eq:PPF_S0}). Using the expressions of $\Delta_{de}$ and $\Theta_{de}$ in eqs.~(\ref{eq:PPF_drho_de}) and (\ref{eq:PPF_theta_de}), we have 
\begin{equation}
  S=S_0+ { aC_1\over \rho_{de}}\Gamma+S_1(\Gamma'+\mathcal{H}\Gamma-S),\label{eq:PPF_S_node}
\end{equation}
where
\begin{equation}
  S_1=  { 3\mathcal{H}aC_1\over k^2 c_KF\rho_{de}}-\frac{a(C_3+3\mathcal{H}D_1)(\rho_T + p_T)+a^2QD_1}{k^2 c_KF\rho_{de}( 1+ w)(\rho_T + p_T)}.\nonumber
\end{equation}

\acknowledgments
  We acknowledge the use of {\tt CosmoMC}. 
  This work was supported by the National SKA Program of China (Grants Nos. 2022SKA0110200 and 2022SKA0110203) and the National Natural Science Foundation of China (Grants Nos. 11975072, 11835009, and 11805031).

\bibliographystyle{JHEP.bst}
\bibliography{idecamb}

\end{document}